\documentstyle[aps,prl,twocolumn]{revtex}
\input epsf

\draft

\begin{document}

\title{Production Efficiency of Ultracold Feshbach Molecules in Bosonic and Fermionic Systems}
\author{E. Hodby, S. T. Thompson, C. A. Regal, M. Greiner, A. C. Wilson, D. S. Jin, E. A. Cornell, and C. E. Wieman}
\address{JILA, National Institute of Standards and Technology and the University of Colorado, and the Department of Physics, University of Colorado, Boulder, Colorado 80309-0440}
\date{\today}

\maketitle
\begin{abstract}
We investigate the production efficiency of ultracold molecules in bosonic $^{85}$Rb and fermionic $^{40}$K when the magnetic field is swept across a Feshbach resonance. For adiabatic sweeps of the magnetic field, the conversion efficiency of each species is solely determined by the phase space density of the atomic cloud, in contrast to a number of theoretical predictions. Our novel model for the adiabatic pairing process, developed from general physical principles, accurately predicts the conversion efficiency for {\it both} ultracold gases of bosons and of fermions. In the non-adiabatic regime our measurements of the $^{85}$Rb molecule conversion efficiency follow a Landau Zener model, with a conversion efficiency that is characterized by the density divided by the time derivative of the magnetic field. 
\end{abstract}


The production of ultracold diatomic molecules is an exciting area of research, with applications ranging from the search for the permanent electric dipole moment \cite{Hudson} to providing unique experimental access to the predicted BCS-BEC cross-over physics \cite{Randeria}. A widely used production technique involves the association of ultracold atoms into very weakly bound ($\sim$10 kHz binding energy) diatomic molecules by applying a time varying magnetic field in the vicinity of a Feshbach resonance. By using a slow adiabatic sweep of the field through the resonance into the region where bound molecules exist \cite{Goral,Mies,Abeelen}, samples of over $10^5$ molecules at temperatures of a few tens of nK have been produced from both quantum degenerate two-component Fermi gases\cite{Jinfirst,Hulet,Salomon} and atomic Bose-Einstein condensates \cite{GrimmCs,KetterleNafirst,Rempe}. 
Over the past two years, several experiments have probed the unique and exotic properties of these molecules. In \cite{Wiemanoscillations} the first coherent superposition of two chemically different species, atoms and Feshbach molecules, was demonstrated, and the vanishingly small molecular binding energy was measured. Bose-Einstein condensation (BEC) has been observed in Feshbach molecules formed from two fermionic atoms \cite{JinBEC,Joachim}, and these molecules provide a unique tool for the controlled investigation of the basic pairing phenomena relevant to superconductivity \cite{Regal2}. 

Despite the widespread use of this technique, the molecule production process has received little experimental attention, and a microscopic description that accurately predicts the conversion efficiency as a function of sweep rate, atom type, density and temperature has not yet been developed. Theoretical work on bosonic systems always assumes the existence of a condensate and hence 100$\%$ molecular conversion for sufficiently slow sweeps, although lower values are always observed in experiments due to finite molecular lifetimes. Meanwhile in quantum degenerate fermionic systems, conversion limits of 50$\%$ have been suggested in certain limiting cases \cite{Pazy} but these theories do not cover the full range of experimentally accessible parameters \cite{JinBEC}. In this paper we present conversion data in the adiabatic regime from both an ultracold but non-condensed cloud of bosonic $^{85}$Rb atoms and an ultracold cloud of $^{40}$K fermions, at a range of degeneracies. We show that a {\it single}, very general description of the pairing process accurately predicts the relationship between conversion efficiency and phase space density in both cases. This theory demonstrates that the complete conversion of the condensate is just the limiting behaviour of a bosonic thermal cloud as the phase space density increases. 

Outside the adiabatic regime, several predictions have been made for conversion efficiency as a function of magnetic field sweep rate \cite{Goral} (and refs. therein) but have only been tested against experimental data over a narrow range of parameters \cite{Jinfirst,Hulet}. 
In the first section of this paper, we present a thorough investigation of the non-adiabatic regime using the $^{85}$Rb system. Our results support a general Landau Zener type theory for conversion as a function of magnetic sweep rate and density that is discussed in detail in \cite{Goral,Mies}.

Detailed descriptions of the $^{85}$Rb experimental apparatus and the magnetic field sweeps for producing molecules are contained in \cite{Cornish} and \cite{Thompson}, respectively. We use evaporative cooling to produce thermal clouds of 50,000$-$130,000 atoms at temperatures of 26$-$94$\,$nK. The atoms are held in a purely magnetic `baseball' trap \cite{Cornish}, with fixed trapping frequencies of 17.6 $\times$ 17.6 $\times$ 6.8 Hz. For efficient  evaporation, the bias field is held at 162~G, where the scattering length $a$ is positive. For slow magnetic field ramps, Rb$_2$ molecules are only produced when the field is ramped upward through the resonance, which is located at 155~G; hence the first step in molecule production is to rapidly jump the magnetic field from 162 to 147.5~G. We then sweep the field back up to 162~G at a chosen linear rate, producing molecules as we pass through the Feshbach resonance. The cloud is then held at 162~G for 10~ms, during which time all the molecules undergo spontaneous one-body dissociation and the two dissociated atoms rapidly leave the cloud with several mK of kinetic energy \cite{Thompson,Koehler}. Thus the number of molecules formed is just the difference in atom number before and after the molecular conversion sweep. Given the unusually low densities of the cloud in our experiments ($\sim$10$^{11}~$cm$^{-3}$), two and three-body atomic decay and collisional molecular decay rates are negligible and do not affect our measurement of the molecular fraction \cite{Thompson}. In other experiments, operating at higher densities (e.g. \cite{KetterleNafirst}), collisional decay {\it cannot} be ignored.

\begin{figure}
\begin{center}\mbox{ \epsfxsize 3in\epsfbox{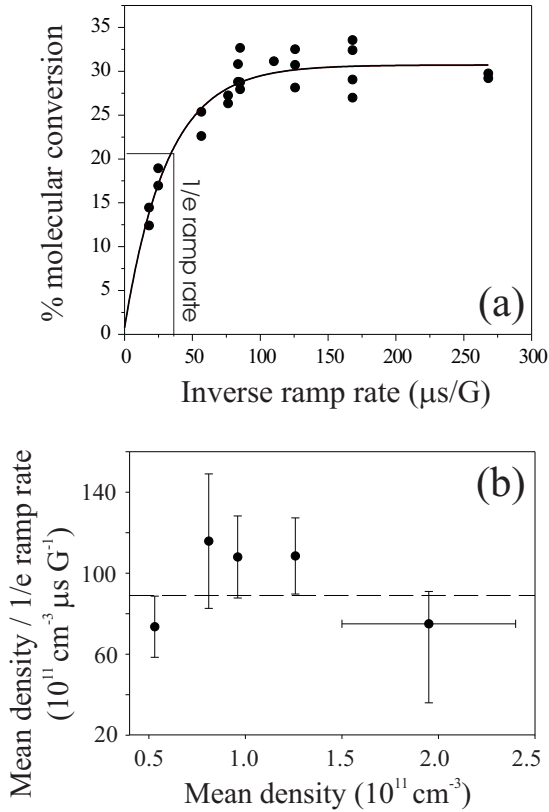}}\end{center}
\caption{(a) Molecular conversion in $^{85}$Rb as a function of inverse magnetic field sweep rate. The initial conditions of the atomic cloud were $N = 100,000$, $T = 30$~nK and $n = 1.5 \times 10^{11}\,$cm$^{-3}$. The solid line is a fit to Eq.(\ref{neqn}). (b) Initial mean density of the atomic cloud ($n$) divided by 1/e sweep rate ($\dot{B}_{1/e}$) and plotted against mean density. The dashed line shows the best fit of Eq.(\ref{deltaeqn}) to the data. For the four low density points, the vertical error bars are dominated by the uncertainty in fitting $\dot{B}_{1/e}$. At higher densities, the cloud experienced significant heating during the ramps across the resonance; hence the density of the final point has significant uncertainty. This heating limited the maximum density used.}\label{nafig} 
\end{figure}

First we investigated the molecular conversion efficiency as a function of magnetic field ramp rate. A typical data set is shown in Fig.~\ref{nafig}(a) and is well fitted by a Landau Zener formula for the transition probability at a two-level crossing  
\begin{equation}
N_{mol} = N_{max}(1-e^{-\beta / \dot{B}}),
 \label{neqn}
\end{equation}
where $N_{max}$ is the asymptotic number of molecules created for a very slow ramp, $\dot{B}$ is the magnetic field sweep rate and $\beta$ is a fitting parameter. $\beta / \dot{B}$ is often called the Landau Zener parameter $\delta_{lz}$.

The dependence of $\delta_{lz}$ on the mean density of the cloud $n$, the magnetic field width of the resonance $\Delta$, the background scattering length $a_{bg}$, and $\dot{B}$ can be derived intuitively up to a constant $\alpha$ by considering the time taken to cross the resonance divided by the mean-field coupling strength,  
\begin{equation}
\delta_{lz} = \alpha \, n \, \Delta \, a_{bg} \, / \, \dot{B}.
 \label{deltaeqn}
\end{equation}

Equation~\ref{deltaeqn} shows that for true Landau Zener behaviour, the quantity $n/\dot{B}_{1/e}$ should be constant, independent of density. (At $\dot{B}_{1/e}$, $\delta_{lz} =1$ and $N_{mol}/N_{max}=63\%$.) Similar datasets to Fig.~\ref{nafig}(a) were taken over a range of initial conditions and $n/\dot{B}_{1/e}$ is displayed versus $n$ in Fig.~\ref{nafig}(b). The data support a constant value for $n/\dot{B}_{1/e}$ and hence a Landau Zener dependence on sweep rate. The value for $\alpha$, extracted from the data of Fig.~\ref{nafig}(b) is $4.5 (4) \times 10^{-7}$ m$^{2}$ s$^{-1}$. In Ref.~\cite{Goral}, the Landau Zener behaviour is rigourously derived for a zero temperature condensate. The authors predict $\alpha$ to be roughly 1/8 of the value that we extract from our thermal cloud data. This difference is currently being investigated \cite{KoehlerPC}.

\begin{figure}
\begin{center}\mbox{ \epsfxsize 3in\epsfbox{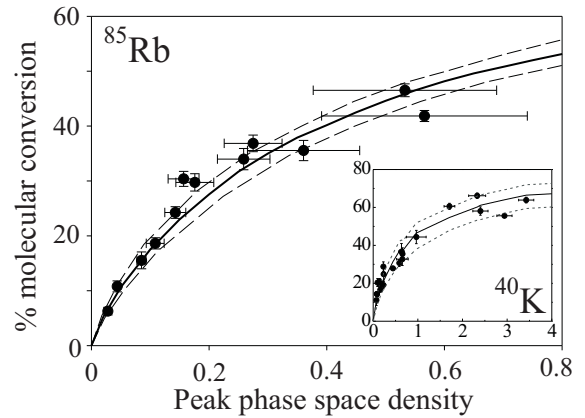}}\end{center}
\caption{Molecule conversion efficiency in $^{85}$Rb as a function of peak phase space density (circles). The solid line shows a simulation based on our conversion theory, fitted to the data with a single pairing parameter $\gamma_b = 0.44 (3)$ (Eq.(\ref{ppeqn})). The dashed lines indicate the uncertainty in $\gamma_b$. Inset: Molecule conversion efficiency in $^{40}$K as a function of peak phase space density (for full data range see Fig.~\ref{fconv}). The best-fit simulation based on our conversion theory is shown as a solid line, with the uncertainty indicated by dotted lines.}\label{bconv} 
\end{figure} 
                           
Our second experiment used both bosonic and fermionic systems and investigated molecular conversion efficiency at a sweep rate that is slow (adiabatic) with respect to $\dot{B}_{1/e}$ (e.g. slower than 150~$\mu$s/G for the conditions of Fig.~\ref{nafig}(a)) but still fast compared to the rate of molecule production via three-body inelastic collisions. This regime \cite{Williams} is important for producing molecules with a limited lifetime. The conversion efficiency for $^{85}$Rb is shown in Fig.~\ref{bconv} as a function of the peak phase space density of the cloud. Note that for adjacent points on Fig.~\ref{bconv} (i.e. points with very similar phase space density), the number of atoms often varies by over a factor of two, indicating that the conversion efficiency is related to phase space density but {\it not} to number or temperature individually. The inset plot in Fig.~\ref{bconv} shows that the same relationship between conversion efficiency and phase space density may be observed in ultracold fermionic $^{40}$K, as discussed later in this paper.

One can argue intuitively that there should be a close relationship between phase space density and conversion efficiency via the adiabatic sweep process. An adiabatic process smoothly alters the wavefunction describing atom pairs but does not change the occupation of states in phase space. Thus to form a molecule, a pair of atoms must initially be sufficiently close in phase space (small relative momentum compared to $h$ divided by separation) that their combined wavefunction can evolve smoothly into the highest bound molecular state as the resonance is crossed. This theory predicts $100 \%$ conversion in a condensate because every atom occupies the {\it same} state in phase space and so is able to form a molecule with any other atom. For a non-condensed sample, we are still able to convert a substantial fraction to molecules because there are no constraints on the center of mass velocity of the atom pair; it simply converts to the velocity of the molecule.

An analytical relationship between phase space density and conversion efficiency is only possible in the limit where the conversion fraction is very small. For almost any finite efficiency, one cannot ignore that the formation of molecules reduces the pairing options for those atoms that remain, leading to the saturation behaviour observed in the experimental data of Fig.~\ref{bconv}. To account for this, we must add a monogamy clause to our model; two atoms will form a molecule if they satisfy the phase space proximity condition described above {\it and} neither has already paired with another atom to form a molecule. To test this model for the pairing process we used a Monte Carlo type simulation based on these two principles. For each run, we randomly assigned positions and momenta to typically 10,000 atoms \cite{numbernote} ensuring that the ensemble retained the correct phase space distribution for the specified temperature and trap frequencies. The simulation program searched for a partner for each atom in turn, removing both atoms from future searches if the pairing condition was satisfied. The simulation ended when no more molecules could be formed. Atoms were considered sufficiently close in phase space to form a molecule if they satisfied the relation
\begin{equation}
| \delta r_{rel} \, m \, \delta v_{rel} | < \gamma \,h, 
\label{ppeqn}
\end{equation}
where $\delta r_{rel}$ is the separation of the pair, $m$ is the atomic mass, $\delta v_{rel}$ is the relative velocity between the two atoms, and $\gamma$ is a constant to be determined by fitting the output of the simulation to experimental data.
Finally, for the boson case, the simulated conversion fraction was increased by 2$\%$ to allow for two-body correlations that assist molecule formation in a cloud of identical bosons \cite{corrnote}.

For each value of $\gamma$, the simulation was run over a range of phase space densities and compared to the experimental data using a least squares routine. The best fit for the bosonic data was given by $\gamma_b = 0.44 (3)$ and is shown as the solid curve on Fig.~\ref{bconv}. The error on this result is dominated by a $12\%$ systematic uncertainty in the phase space density (resulting from uncertainties in atom number and temperature). The dotted lines indicate the range of values for the pairing parameter $\gamma_b$ that result from this phase space density uncertainty. The excellent agreement between experiment and simulation over a wide range of phase space density provides strong evidence in support of our molecular conversion model.

Our model for the pairing process is based on very general arguments about adiabaticity and atom availability. With appropriate phase space distributions, it is equally applicable to molecule formation in a fermionic system. This provides another stringent test of its validity. To study the molecule conversion efficiency in a Fermi system, we
create an ultracold, two-component Fermi gas as outlined in \cite{Regal1}. To achieve a wide range of phase space density, we follow
cooling of the gas in an optical dipole trap with a recompression
of the trap and controlled parametric heating through a modulation
of the trap strength. This allows creation of Fermi gases ranging
in temperature from $<0.05 \, T_F$ to 1.3~$T_F$, where $T_F$ is the
Fermi temperature. The dipole trap that confines the final Fermi
gas is characterized by radial frequencies, $\nu_r$, between 312
Hz and 630 Hz and an aspect ratio of $\nu_r/\nu_z =70 \pm 20$. With this ultracold
Fermi gas, we create molecules as described in \cite{Jinfirst} using a broad
s-wave Feshbach resonance between the $|f,m_f \rangle = |9/2, -9/2 \rangle $
and $|9/2,-7/2 \rangle$ states of $^{40}$K located at
$B_0=202.1 \pm 0.1$ G \cite{Regal2}. The magnetic field sweep used to create
molecules starts at 202.83 G (where $a<0$), ends between 201.10 G and 201.59
G and occurs at inverse sweep speeds ranging from 640 to 2900 $\mu
s$/G. Within this range the conversion fraction is independent of sweep rate, confirming that the conversion is adiabatic. On the time scale of these magnetic field sweeps, the loss of
molecules and atoms is negligible \cite{Regal3}. At the end of the magnetic
field sweep, we immediately turn off the dipole trap and allow the
gas to ballistically expand. We then use one of two techniques to
probe the molecular conversion: (1) We compare the number of
atoms remaining on the $a >0$ side of the resonance, to the number of atoms measured at $a < 0$
as described in \cite{Jinfirst}. (2) We convert the
remaining atoms in the $|9/2,-7/2 \rangle$ state to the
$|9/2,-5/2 \rangle$ state by applying a radio frequency $\pi$
pulse.  We then dissociate the molecules by ramping the magnetic
field back to the $a<0$ side of the Feshbach resonance. Finally,
we image the $|9/2,-7/2 \rangle$ and $|9/2,-5/2
\rangle$ states separately to extract the molecule and atom
numbers, respectively.

\begin{figure}
\begin{center}\mbox{ \epsfxsize 3in\epsfbox{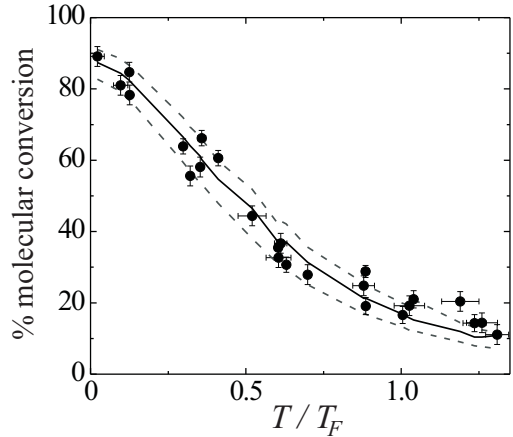}}\end{center}
\caption{Molecule conversion efficiency in $^{40}$K as a function of $T/T_F$. (Note that $T/T_F$ uniquely determines the phase space density of a fermionic cloud). The initial clouds have mean densities for each spin state ranging from $1 \times 10^{12}$ to $2 \times 10^{13}$cm$^{-3}$ and 52(2)$\%$ of the atoms are in the m$_f$ = -7/2 state. The data are fitted with the same conversion model as for $^{85}$Rb. The best fit curve has a pairing parameter of 0.38(4). The dotted lines indicate the uncertainty on this result. Note that conversion efficiencies far greater than $50 \%$ are measured.}\label{fconv} 
\end{figure}

Experimental data for molecule formation efficiency in the fermionic gas is shown in Fig.~\ref{fconv}, and a limited range of the same data is shown in the inset of Fig.~\ref{bconv}. The data was simulated with the {\it same} molecular conversion model as the bosonic data, modified to provide the correct Fermi-Dirac phase space distributions for a 50/50 spin mixture and allow pairing only between atoms with unlike spins. The pairing parameter for the best fit simulation is $\gamma_f = 0.38 (4)$, where the error includes a $10 \%$ systematic uncertainity in $T/T_F$. This result is in excellent agreement with the value of $\gamma_b = 0.44 (3)$ from the bosonic system, especially given that both results are dominated by systematic rather than statistical uncertainties. This agreement provides clear evidence that our model and the physics underlying it accurately describes molecular conversion efficiency in both bosonic and fermionic systems. 
One fundamental difference between our model and many previous theories is that we allow each atom to pair with {\it any} other available atom that satisfies the phase space proximity condition. Other theories have considered only one potential partner for each atom, which has led to a ceiling for the fermionic conversion of 50$\%$ \cite{Pazy}, contradicting the experimental data of Fig.~\ref{fconv}. 

In summary, we have performed the first thorough experimental investigation of the efficiency of molecular formation when the magnetic field is swept across a Feshbach resonance. In the non-adiabatic regime, we have shown that the conversion efficiency follows a dependence on density and sweep rate that is well approximated by a two level Landau Zener model. In the adiabatic regime, using an ultracold but non-condensed gas of bosonic $^{85}$Rb atoms and a quantum degenerate gas of fermionic $^{40}$K, we find that the conversion efficiency is monotonically related to the initial phase space density of the atomic cloud. We have shown that the pairing process allows each atom to form a molecule with any other atom in the cloud provided that they are sufficiently close together in phase space. The agreement of experimental data from both bosonic and fermionic systems with a single simulation provides compelling evidence for the validity of this molecule conversion model. 

We thank J. Stewart for experimental assistance. This work has been supported by ONR, NSF and NIST. S.T.T. acknowledges support from ARO-MURI and C.A.R. from the Hertz Foundation.


\begin{references}
\bibitem{corrnote}
The factor of $2 \%$ was determined by adding atoms to a Gaussian distribution at appropriate locations to approximate a Bose distribution with the correct two-body correlations. The conversion efficiency of the correlated and original Gaussian distributions could then be compared. It is only 2 $\%$ because of the small fraction of pairs within a thermal de Broglie wavelength.
\end{references}
\end{document}